\documentclass[twocolumn,showpacs,aps,floatfix]{revtex4}

\usepackage{amsmath}
\usepackage{graphicx}
\usepackage{dcolumn}
\usepackage{bm}

\begin{document}

\title{A simple method for obtaining electron scattering phase shifts from energies of an atom in a cavity}
\author{I.M. Savukov}

\affiliation{Department of Physics, Princeton University,
Princeton, New Jersey 08544}

\date{\today}

\begin{abstract}
We present a simple method for obtaining elastic scattering phase
shifts and cross sections from energies of atoms or ions in
cavities. This method does not require calculations of
wavefunctions of continuum states, is very general, and is
extremely convenient from practical point of view: some
conventional computer codes designed for the energies of bound
states can be used without modifications. The application of the
method is illustrated on an example of electron scattering from Kr
and Ar. From Brueckner orbital energies in variable cavities, we
have obtained ab initio cross sections that are in close agreement
with experiment. The relativistic effects are also considered and
found to be small below 10 eV.
\end{abstract}
 \maketitle


Conventional methods of calculations of scattering cross sections
are cumbersome, inconvenient, and very often inaccurate. This is
only because they all are based on computing continuum, or
sometimes quasicontinuum, wavefunctions and asymptotic fittings to
extract phase shifts. Such an approach requires modifications of
conventional atomic structure codes, developed for bound states,
or just writing new programs altogether. For a known potential it
is not a difficult task -- this is why numerous semi-empirical
calculations can be found in the literature -- but the level of
accuracy and theoretical uncertainty of calculations based on ad
hoc potentials can not be totally satisfactory. For ab initio
calculations already complicated codes have to be rewritten, which
takes considerable amount of time. For multi-configuration
Hartree-Fock (MCHF) method this was undertaken by
Saha~\cite{SahaNeon,SahaArgon} to obtain ab initio results in
agreement with experiment. However, many-body perturbation theory
(MBPT) methods, which were developed for fundamental symmetry
tests, have not been used for calculations of electron scattering
cross sections.

The method we propose in this letter is very simple and general:
instead of finding continuum wavefunctions and fitting them to
asymptotical solutions to obtain phase shifts for given electron
energies, we impose a boundary condition on an atom, an ion, or a
molecule to make the spectrum discreet and then from discreet
energies extract phase shifts which are uniquely related to these
energies. Thus the problem of phase shifts is converted into a
conventional problem of finding energies of bound states.
Especially simple relation exists, as we will show, in the case of
an atom in a spherical cavity.

It can be shown that continuum and quasicontinuum wavefunctions
are equivalent. For example, in Ref.~\cite{bsplinescont} it was
stated that B-spline solutions obtained in a cavity can be
interpreted as a representation of true continuum states with a
different normalization, and the energy of the quasicontinuum
states can be set to an arbitrary positive value by adjusting the
size of the cavity. There are also other methods that give
B-spline continuum wavefunctions at any energy: the Galerkin
method~\cite{Froesescat}, least-squares approach~\cite{Brosolo,
Brosolo2}, and free boundary condition
approach~\cite{Nikolopoulos}. The emphasis in these works is
placed on applications of B-splines which are very often bundled
with the cavity boundary conditions: for the method proposed here,
however, the boundary conditions are more essential than
B-splines, which are still convenient for evaluation of radial
integrals in high-precision MBPT calculations~\cite{PRLspline}.

While our method can be justified mathematically in quite general
assumptions, it is not yet obvious that the method will be
accurate in practical calculations, so we will illustrate the
usefulness and accuracy of the method on specific examples such as
the MBPT calculations of electron elastic scattering phase shifts
and cross sections from Ar and Kr. The MBPT is chosen because it
can provide the best accuracy for a negative mono-valent ion, e.g.
Ar+e$^-$, uses cavity-bound basis functions, so the code does not
need modification, and allows systematic consideration of
correlations. To obtain correct electron-noble gas scattering
cross sections it is necessary to include the direct and exchange
potentials from a frozen noble-gas atom as well as core
polarization effects. The direct part of the unperturbed atomic
potential produces  a phase shift opposite to that of the exchange
part. The combined phase shift from the frozen atom is opposite in
sign to that of the core-polarization. We will show that so-called
Brueckner-orbital (BO) approximation treats accurately these
effects and results in good precision. The accuracy can be further
improved by using all-order couple-cluster method, or other
accurate methods developed for monovalent atoms. The calculations
of phase shifts from energies for other systems should be also
possible and will be undertaken in future.

Apart from illustration purpose, the calculations will serve to
provide accurate ab initio cross sections for comparison with
other theories and experiments and to improve understanding of
this particular system. Despite longer than a century
history~\cite{rivista} that experiments on electron interaction
with gases have, many questions remain open and this area of
research is still very active. Elastic scattering of electrons on
noble-gas atoms is of particular interest since many precise
measurements are available providing tests for theories which all
with a few exceptions are not of ab initio type and are based on
pseudopotentials to take into account exchange interaction and
significant polarizability of noble-gas atoms by an electron.
Although elaborate complicated semi-empirical effective potentials
have been developed to achieve good accuracy of calculations, many
different calculations and measurements are still in disagreement,
and there is clearly significant uncertainty in theoretical
understanding. This situation exist in almost all noble-gas atoms.

For example, motivated by uncertainty in cross sections at low
energies, which are important for extraction of scattering
lengths, first ab initio calculations of low-energy electron
scattering from neon~\cite{SahaNeon} and argon~\cite{SahaArgon}
based on MCHF method to account for polarization effects have been
reported, and good agreement with experiment has been
demonstrated. Although MCHF method is very effective in general
for the consideration of complicated open-shell ions, in
monovalent atoms and low-charge ions this method has lower
accuracy than MBPT methods, which were not applied to calculations
of electron scattering from noble-gas atoms, probably due to
complications associated with continuum states.

With the aid of partial wave expansion,
\begin{equation}
\Psi(\textbf{r})=\sum_{lm}Y_l^m(\theta,\phi) \frac{P_l(r)}{r}
\end{equation}
 a total elastic cross section $\sigma _{t}$ can be
found from phase shifts $\delta _{l}$
\begin{equation}
\sigma _{t}=\frac{4\pi }{k^{2}}\sum (2l+1)\sin ^{2}\delta _{l}
\end{equation}
which are normally extracted from asymptotic behavior of radial
wavefunctions $P_{l}(r)$ obtained by numerical solution of radial
Schr\"{o}dinger equation
\begin{equation}
\frac{d^{2}P_{l}(r)}{dr^{2}}+\left[
k^2-U(r)-\frac{l(l+1)}{r^{2}}\right] P_{l}(r)=0
\end{equation}
for a given energy $E=k^2/2$ as a parameter. In this equation
$U(r)$ is some effective potential which describes approximately
direct and exchange interaction as well as the attraction due to
core polarizability. (Atomic units are used in all equations.) The
radial wavefunctions can be also obtained by using ab intio atomic
structure methods such as MCHF~\cite{SahaArgon, SahaNeon} or MBPT.
Because wavefunctions are not always available in precision MBPT
calculations and most codes output either energies or matrix
elements, the extraction of phase shifts from wavefunctions is not
very convenient. However, it is not necessary: phase shifts can be
obtained from energies of an atom bound to a cavity, which is a
natural setting in MBPT calculations. The extraction is possible
because the cavity uniquely encodes phase shift information into
energies of quasicontinuum states and quasicontinuum wavefunctions
are proportional to true continuum wavefunctions if their energies
are the same. The last statement can be easily proved since the
continuum and quasicontinuum wavefunctions are both unique
solutions of the radial differential equation with the same
boundary condition at $r\rightarrow 0$, the same energies,
although with different normalization conditions and maybe sign
convention. The equivalence of quasicontinuum and continuum states
was also stated in Ref.~\cite{bsplinescont}. At large $r$
continuum and quasicontinuum solutions approach asymptotically the
solution in empty cavity proportional to $rj_{l}(r)$, where
$j_{l}(r)$ are spherical Bessel functions, and the effect of the
atomic potential is only in phase shifts which can be determined
from the asymptotic form of the wavefunctions or from energies for
a known cavity radius $R$, which is our proposed method:
\begin{equation}
\delta_l(E_n) =x_{ln}-\sqrt{2E_n}R \label{zeros}
\end{equation}
where $x_{ln}$ is the n$^{th}$ zero of the spherical Bessel
function $j_l(x)$. Accurate values of $x_{ln}$ can be found in
mathematical reference books, for example on page 467 of
Ref.\cite{HMF}, column $j_{\nu,s}$ where $\nu=l+1/2$ and $s=n$.
For $l=0$, $x_{0n}=n\pi$. The lowest quasicontinuum state of a
given symmetry has to be used with the first zero of the
corresponding spherical Bessel function, the next state with the
second zero, etc.

Energies of quasicontinuum states are calculated in
Brueckner-orbital approximation, which accounts for
core-polarization effects with relatively high precision. First,
the Dirack-Hartree-Fock (DHF) equation is solved for a
closed-shell atom (Ar or Kr). Then in the obtained DHF potential,
B-spline finite basis is generated. In this basis, the Hamiltonian
matrix $h_{ij}=\delta_{ij}\epsilon_i+\Sigma _{ij}(\varepsilon
_{0})$,
\begin{eqnarray}
\Sigma _{ij}(\varepsilon _{0}) &=&\sum_{kcmn}\frac{%
(-1)^{j_{m}+j_{n}-j_{i}-j_{c}}}{(2j_{i}+1)(2k+1)}\frac{X_{k}(icmn)Z_{k}(mnjc)%
}{\varepsilon _{0}+\varepsilon _{c}-\varepsilon _{m}-\varepsilon
_{n}}+
\nonumber \\
&&\sum_{kbcn}\frac{(-1)^{j_{i}+j_{n}-j_{b}-j_{c}}}{(2j_{i}+1)(2k+1)}\frac{%
X_{k}(icmn)Z_{k}(mnjc)}{\varepsilon _{0}+\varepsilon
_{n}-\varepsilon _{b}-\varepsilon _{c}} \label{SelfEnergy}
\end{eqnarray}
is calculated and diagonalized to obtain BO energies. The
summation runs over core states $c$, excited states $n,m$, and
angular momenta $k$; the matrix elements are calculated between
all possible states $i$ and $j$. The coupled radial integrals
$X_{k}(abcd)$ and $Z_{k}(abcd)$ are defined for example
in~\cite{Zdefine}. The self-energy matrix elements $\Sigma
_{ij}(\varepsilon _{0})$, which take into account dominant part of
core-polarization effects, depend on electron energy $\varepsilon
_{0}$ and contains non-local interaction, so that they can not be
approximated accurately with a single effective potential unless
the energy range is small, $\varepsilon\leq\varepsilon _{0}$, and
the distance between electron and an atom is large compared to the
size of the atom so that exchange interaction can be neglected.
The diagonalization is important because energy differences
between quasi-continuum states are small. Essentially, all-order
methods are necessary, at least to include chained self-energy
corrections. Pure 2nd- or 3rd-order expansions will be inaccurate
due to this reason, and we will illustrate this numerically for
2nd-order MBPT in the next section, but couple-cluster methods,
which treat some diagrams in all orders, are expected to give good
accuracy. The simplest future improvement for the current BO
theory is to take into account screening, which is more
significant in heavier noble-gas atoms.

Relativistic effects can be also carefully considered, if
necessary. One
effect is the difference in energies between for example $p_{1/2}$ and $%
p_{3/2}$ states, which for low-energy scattering is small, but
becomes more pronounced at higher energies. The self-energy
correction is also slightly different in non-relativistic and
relativistic cases because intermediate states in the summation
are different. One interesting consequence of the use of
relativistic basis in calculations is that the boundary condition
is not $P(R)=0$, but rather $P(R)=Q(R)$, where $P(r)$ and $Q(r)$
are large and small components of the radial Dirac wavefunction.
This is so-call ``bag'' boundary condition which is required to
avoid Klein paradox~\cite{Kleinpardox} and spurious solutions
observed in Ref.~\cite{spur}. Using the Pauli expansion, it can be
shown that the difference in boundary
conditions produces additional phase shift equal to $\alpha \sqrt{E/2}$, where $%
\alpha $ is the fine-structure constant and $E$ is the energy of
the electron. This shift can be obtained if we compare energies
generated in the empty cavity with energies expected from the
zeros of the spherical Bessel functions as illustrated in
Table~\ref{table2}.
\begin{table}
\caption{``Bag'' model artifact.  An extra phase shift $\delta$
due to the ``bag'' boundary condition for an empty cavity of R=15
a.u. is compared with prediction $\alpha \sqrt{E/2}$ in the Pauli
approximation; $l$ is the angular momentum of the state, $n$ is
the radial quantum number.} \label{table2}
\begin{tabular}{lllll}
\hline\hline
$l$ & $n$ & $E_{cav}$ & $\delta$  & $\alpha \sqrt{E/2}$ \\
\hline
1 & 1 & 4.48[-2] & 1.10[-3] & 1.09[-3] \\
1 & 3 & 2.64[-1] & 2.69[-3] & 2.65[-3] \\
1 & 4 & 4.39[-1] & 3.80[-3] & 3.42[-3] \\
2 & 1 & 7.38[-2] & 1.41[-3] & 1.40[-3] \\
\hline
\end{tabular}
\end{table}
Apparently, the Pauli approximation explains well and predicts
accurately the ``bag'' shift as long as $\alpha \sqrt{E/2}\ll 1$.
When the shift is large, it is necessary to subtract it or even to
reanalyze this method more carefully. In the calculations
presented below the energies were small enough to neglect this
effect as well as some other relativistic effects.

\begin{figure}
\centerline{\includegraphics*[scale=0.8]{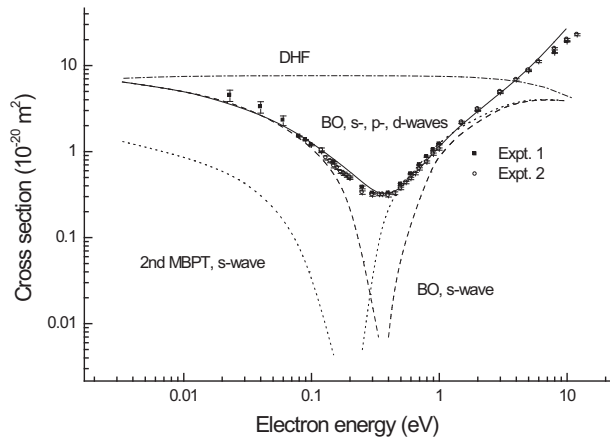}}
\caption{Low energy argon cross section. Comparison of various
theoretical approximations and of the final accurate BO values
with experiment. The solid line (partial waves l=0-2) and the
dashed line (l=0) show our BO cross sections obtained after
diagonalization of the Hamiltonian matrix that contains
self-energy defined by Eq.(\ref{SelfEnergy}). The dotted line is
cross section obtained from 2nd order MBPT energies without
diagonalization. The dash-dotted line shows cross section obtained
from DHF energies. Experimental results are taken from
Refs.~\cite{Gushkov} (Expt. 1) and \cite{BuckmanLohmann}
(Expt.2).} \label{argoncross}
\end{figure}
\begin{figure}
\centerline{\includegraphics*[scale=0.7]{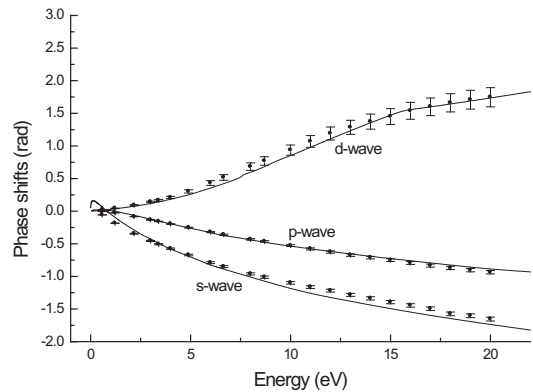}}
\caption{Comparison of our BO calculations (the solid line) with
experiment \cite{Williams} (the points with error bars) for phase
shifts of electron scattering from argon.} \label{phase}
\end{figure}
The results of our calculations for elastic cross section on argon
is shown in Fig.\ref{argoncross}. Close agreement with
experimental data is achieved in the range below 10 eV if the
self-energy chain corrections are included (BO energies are used)
and the cross sections from partial waves with $l=0-2$ are added.
To emphasize the importance of $l>0$ contributions in
Fig.\ref{argoncross} we also plot $s$-wave cross section
separately and in Fig.\ref{phase} we compare phase shifts from
$s$-, $p$-, and $d$-waves. To check that our predictions for phase
shifts are correct, we compare them with experimental phase
shifts. Contributions from higher order partial waves are much
smaller, but can be in principle included. At low energies, the
dominant contribution comes from $s$-waves, which is expected;
however, at energy about 0.36 eV, $s$-wave $\sin\delta$ crosses
zero, resulting in a minimum of the cross section. In this region
the $p$-wave and $d$-wave contributions become particularly
important and affect the shape of the Ramsauer-Townsend minimum.

Although the diagonalization does not change much energies of the
quasi-continuum states, the phase shifts and cross sections
obtained from energies before (dotted line) and after (solid line)
diagonalization  are quite different, see Fig.\ref{argoncross}.
The agreement is achieved only in the last case. In the case when
argon cross section is calculated from DHF energies and thus
polarization effects are ignored, the cross section is completely
inaccurate. At low energy DHF scattering length is exactly
opposite to correct value. DHF potential cross section does not
depend much on energy and this potential in the range
below 3 eV can be approximated by an infinite potential at $R<R_{0}$, where $%
R_{0}=$1.42-1.53 a.u. approximately equal to the size of the argon
electron cloud about 1.56 a.u. Simple interpretation of this is
the repulsion due to Pauli exclusion principle.

Because the experiments at very low energy are difficult, we also
find scattering length by extrapolating our results to zero energy, $%
R_{scatt}=-1.47\pm 0.03$ a.u. There are several other calculations
of the argon scattering length: -1.63 a.u. by~\citet{Asaf} from
studies of perturbed optical absorption in gases , -1.492
by~\citet{BuckmanLohmann} and -1.449 by~\citet{Ferch} from TSC
studies, and -1.486 by~\citet{SahaArgon} from low energy
calculations.
 Our value disagrees only with the value from Ref.\cite{Asaf}.

\begin{figure}
\centerline{\includegraphics*[scale=0.7]{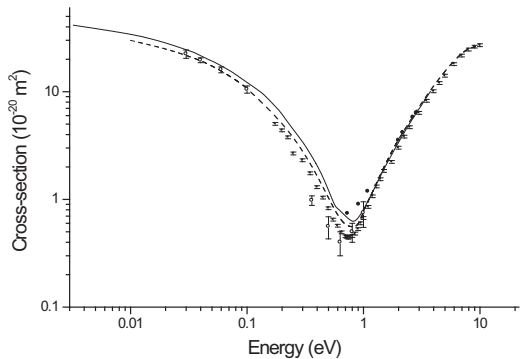}}
\caption{Low-energy krypton cross section. The solid line shows
our theory and dashed line shows pseudopotential calculations
by~\citet{Plenkiewicz}. Experiment: open circles with larger error
bars show the cross sections of~\citet{Gushkov}, solid circles
with smaller error bars show cross sections of~\citet{Buckman},
and solid circles without error bars show results
of~\citet{Subramanian}. } \label{kryptscat}
\end{figure}

Our theoretical cross section for krypton is shown in
Fig.\ref{kryptscat}. The agreement with the cross section obtained
by~\citet{Plenkiewicz} from a pseudopotetial is very close. The
agreement with experiment is also relatively good in all range of
energies shown, although some disagreement can be seen near the
cross-section minimum, which can be due to the inaccuracy of both
theory and experiment.

In this paper, we proposed a simple method for calculations of
phase shifts from energies of quasicontinuum states and
illustrated its high precision with MBPT calculations. The method
in general can be applied to many scattering problems: electron
scattering on various atoms and ions, positron scattering,
atom-atom scattering; however, in each case some specific atomic
structure method has to be developed to achieve practical
precision. Discussed BO approximation can be used only for
electron scattering on closed-shell atoms and ions.

The author is grateful to Dr. Kuzma for finding relevant
references and to Prof. Happer for discussion of physics of
electron scattering, for reading the manuscript, and for
suggestions for its improvement.

\end{document}